\documentclass[prl,twocolumn]{revtex4-2}
\usepackage{graphicx}
\usepackage{amsmath,amsfonts,mathtools}
\usepackage{bm}
\usepackage{hyperref}
\usepackage{physics}
\usepackage{color}
\newcommand{\rev}[1]{{\color{black}#1}}
\begin{document}
\title{Dualities and Topological Classification of the $S=1$ Pyrochlore Spin Ice}
\author{Sena Watanabe}
\email{watanabe-sena397@g.ecc.u-tokyo.ac.jp}
\affiliation{Department of Applied Physics, The University of Tokyo, Tokyo 113-8656, Japan}

\author{Yukitoshi Motome}
\email{motome@ap.t.u-tokyo.ac.jp}
\affiliation{Department of Applied Physics, The University of Tokyo, Tokyo 113-8656, Japan}
 
\author{Haruki Watanabe}
\email{hwatanabe@ust.hk}
\affiliation{Department of Physics, Hong Kong University of Science and Technology, Clear Water Bay, Hong Kong, China}
\affiliation{Institute for Advanced Study, Hong Kong University of Science and Technology, Clear Water Bay, Hong Kong, China}
\affiliation{Center for Theoretical Condensed Matter Physics, Hong Kong University of Science and Technology, Clear Water Bay, Hong Kong, China}
\affiliation{Department of Applied Physics, The University of Tokyo, Tokyo 113-8656, Japan}

\date{\today}

\begin{abstract}
\rev{We derive dual descriptions of the trivial paramagnetic, $\mathbb{Z}_2$-deconfined, and $\mathbb{Z}_2$-confined U(1) Coulomb phases of the $S=1$ pyrochlore spin ice.} 
In the monopole-free limit, the system can be effectively mapped onto 3D $XY$ and Ising loop-gas models depending on the spin anisotropy, which provides theoretical estimates for the phase boundaries, while a macroscopic flux vector classifies the topological sectors via geometric parity rules. 
\rev{Thermal unit monopoles permit endpoints of both directed spin loops and unoriented vacancy loops. The resulting relevant symmetry-breaking fields in the dual $XY$ and Ising descriptions predict rounded crossovers near the monopole-free critical points. Monte Carlo simulations support this monopole-induced rounding.}
\end{abstract}

\maketitle

\paragraph{\textcolor{black}{Introduction.---}}
\rev{Frustrated magnets host phases beyond conventional symmetry-breaking paradigms~\cite{Balents2010, Savary2016, Knolle2019, Broholm2020}. In classical spin-$1/2$ pyrochlore ice~\cite{Harris1997, Ramirez1999, Bramwell2001, Gingras2014, Hermele2004, Isakov2004, Henley2005}, the two-in, two-out rule imposes a coarse-grained divergence-free flux, $\nabla\cdot\bm B=0$, with a Maxwell action~\cite{Huse2003, Moessner2003, Hermele2004}. Its Coulomb phase has dipolar correlations and pinch points~\cite{Henley2005, Isakov2004, Fennell2009, Morris2009}. At finite temperature, mobile monopoles violate the ice rule~\cite{Castelnovo2008, Jaubert2009}, screening these correlations and degrading the Coulomb phase into a paramagnet~\cite{Ryzhkin2005, Castelnovo2011}.}

While the spin-$1/2$ spin ice is well understood, releasing the rigid spin-length constraint opens a frontier for emergent phenomena, including magnetic moment fragmentation~\cite{BrooksBartlett2014, Petit2016}, fractionalized spin liquids~\cite{Rehn2017}, and exotic spin nematics~\cite{PhysRevB.81.184409, PhysRevB.94.174417}. Very recently, Pandey and Damle investigated a spin-$1$ extension of the pyrochlore antiferromagnetic Ising model~\cite{Pandey2025}, building on related developments in bipartite dimer-loop models~\cite{Kundu2025PRX, Kundu2025arXiv}. 
In this model, the spins can additionally occupy a nonmagnetic $S^z=0$ state, whose relative statistical weight is controlled by a fugacity parameter $w$ determined by the single-ion anisotropy.
Through large-scale Monte Carlo (MC) simulations, they unveiled a phase diagram characterized by three distinct regimes: a trivial paramagnetic phase for $w < w_{c_1}$, a $\mathbb{Z}_2$-deconfined U(1) Coulomb phase for $w_{c_1} < w < w_{c_2}$, and a $\mathbb{Z}_2$-confined U(1) Coulomb phase for $w > w_{c_2}$. \rev{Pandey and Damle supported 3D $XY$ and Ising criticality through finite-size scaling and qualitative arguments based on the divergence-free current representation and the dual ferromagnet--nematic transition~\cite{Pandey2025}.} The nature of such topological phases and confinement transitions has long been a central theme in condensed matter physics and lattice gauge theories~\cite{Fradkin1979, Kogut1979, Senthil2000, Kitaev2006}, with implications extending even to water ice~\cite{Benton2016}, hydrogen-bonded ferroelectrics~\cite{Chern2014}, and phononic band insulators~\cite{Han2025}. \rev{For this model, however, the constrained phase diagram alone does not establish thermodynamic transitions at finite $J/T$: even dilute monopoles can be relevant perturbations near criticality. An explicit construction of the statistical weights and fields is needed to determine whether the limiting transitions survive or become crossovers when the constraint is energetic rather than exact.}

\rev{In this Letter, we describe the two constrained transitions through complementary loop gases on the diamond lattice: directed $S^z=\pm1$ loops in an $S^z=0$ background at small $w$, and unoriented $S^z=0$ loops in an $S^z=\pm1$ ice background at large $w$. Explicit dual statistical weights~\cite{Wegner1971}, using a small-$w$ expansion and a large-$w$ Pauling estimate, connect their proliferation to $XY$ and Ising transitions. The central issue is whether these loops can break. At $v=0$ both graphs must remain closed; thermal unit monopoles permit endpoints in both descriptions. These endpoints act as relevant dual ordering fields, rounding the constrained proliferation transitions into crossovers. We also perform direct finite-temperature MC simulations of the original spin model without imposing the monopole-free approximation. The dual effective theories provide a microscopic explanation for the rounded specific-heat peaks and loss of flux quantization observed in these simulations.}

\paragraph{\textcolor{black}{Model.---}}
\begin{figure}[t]
    \centering
    \includegraphics[width=\linewidth]{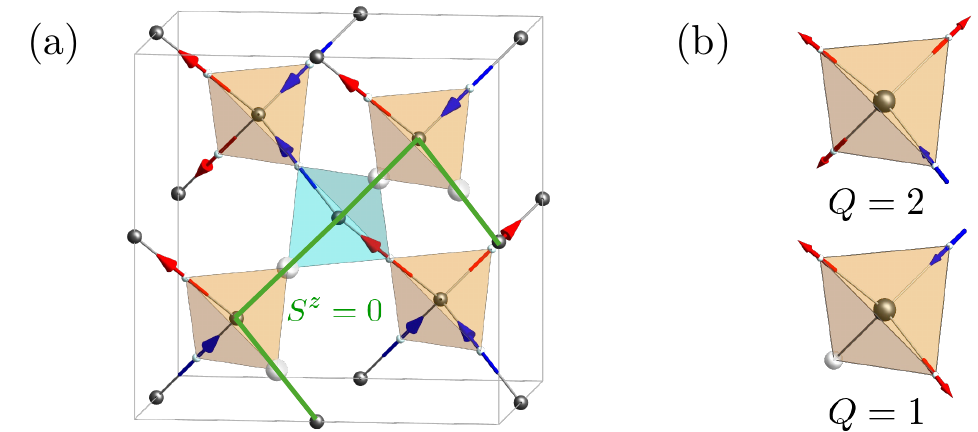}
    \caption{(a) Geometric correspondence between the diamond and pyrochlore lattices. 
    Black spheres represent the diamond sites, which are located at the centers of the cyan and orange tetrahedra formed by the pyrochlore sites.
    Arrows on the pyrochlore sites denote the spin-$1$ variables aligned with the local $z$-axis.
    Red and blue arrows represent $S_\ell^z=1$ and $S_\ell^z=-1$, respectively, while white spheres correspond to $S_\ell^z=0$.
    The green line traces a closed loop of $S^z=0$ links (white spheres)---\rev{an unoriented vacancy loop in the large-$w$ description introduced below.}
    The illustrated spin configuration satisfies the ice rule.
    (b) Monopole charges and their corresponding spin configurations. Odd monopole charges can exist in the spin-1 model.
    }
    \label{fig:diamond_pyrochlore}
\end{figure}

We consider a classical spin-$1$ model of the Blume-Capel type~\cite{Blume1966, Capel1966} on the pyrochlore lattice.
Its sites reside at the midpoints of the nearest-neighbor links $\ell = \langle \bm{r}_A, \bm{r}_B \rangle$ of the underlying bipartite diamond lattice. 
The local positive $z$-axis for each spin $S^z_\ell \in \{-1, 0, 1\}$ points from $\bm{r}_A$ to $\bm{r}_B$. 
This geometry is illustrated in Fig.~\ref{fig:diamond_pyrochlore}.
\rev{These are three classical Ising-type states along local $\langle111\rangle$ axes, with no transverse matrix elements. A crystal-field realization requires a magnetic pair and a nonmagnetic state; projected states need not form an actual spin-$1$ multiplet~\cite{Gingras2014}.}
The system consists of $L^3$ cubic unit cells, each of linear size $1$, subject to periodic boundary conditions. Each unit cell contains $8$ diamond-lattice sites and $16$ pyrochlore sites, totaling $N \coloneqq 8L^3$ diamond-lattice sites and $2N$ spins.

The Hamiltonian $H \coloneqq J\sum_{\langle \ell, \ell' \rangle} S_\ell^z S_{\ell'}^z + \Delta \sum_\ell (S_\ell^z)^2$, comprising a nearest-neighbor exchange $J$ and a single-ion anisotropy $\Delta \coloneqq J + \mu$, can be recast into a perfect-square
\begin{equation}
    H = \frac{J}{2}\sum_{\bm{r}} \Big(\sum_{\ell \in \bm{r}} S^z_\ell\Big)^2 + \mu \sum_\ell (S_\ell^z)^2,\label{eq:model}
\end{equation}
where $\ell\in \bm{r}$ denotes the four links connected to the diamond-lattice site $\bm{r}$. 
Magnetic monopoles are characterized by the discrete divergence of the spin configuration, which is defined for sublattices A and B as
$\nabla \cdot \bm{S}_{\bm{r}_A} \coloneqq \sum_{\ell \in \bm{r}_A} S^z_\ell$ and $\nabla \cdot \bm{S}_{\bm{r}_B} \coloneqq -\sum_{\ell \in \bm{r}_B} S^z_\ell$,
respectively.

Introducing the fugacities $w \coloneqq e^{-\mu/T}$ and $v \coloneqq e^{-J/(2T)}$, the partition function $Z \coloneqq \sum_{\{S_\ell^z\}} e^{-H/T}$ is rewritten as   
\begin{equation}
    Z = \sum_{\{Q_{\bm{r}}\}}\sum_{\{S_\ell^z\}} \left( \prod_{\bm{r}} \delta_{\nabla \cdot \bm{S}_{\bm{r}}, Q_{\bm{r}}} \right) v^{\sum_{\bm{r}}Q_{\bm{r}}^2} w^{\sum_\ell(S_\ell^z)^2}.
\end{equation}
\rev{For $J\gg T$ ($v\ll1$), thermal monopoles are dilute, motivating the strictly constrained reference problem $v=0$~\cite{Pandey2025}. However, a low local monopole density does not imply dominance of the entirely monopole-free sector in thermodynamic volume or justify extending its critical thermodynamics to fixed $v>0$.}

\paragraph{\rev{Directed Spin Loops and the $XY$ Model.---}}
\rev{For small $w$, the nonzero $S^z=\pm1$ links are oriented unit currents in a predominantly $S^z=0$ background. At $v=0$, Gauss's law forces these currents to form closed directed loops, allowing intersections and winding configurations. Their angular representation follows the standard duality of divergence-free current models onto the $XY$ model}~\cite{Jose1977, Savit1980}.
Gauss's law constraint $\nabla \cdot \bm{S}_{\bm{r}} = 0$ is enforced by introducing a continuous phase field $\theta_{\bm{r}} \in [0, 2\pi)$ via the Fourier integral $\delta_{\nabla \cdot \bm{S}_{\bm{r}}, 0} = \int_0^{2\pi} \frac{d\theta_{\bm{r}}}{2\pi} e^{-i \theta_{\bm{r}} \nabla \cdot \bm{S}_{\bm{r}}}$. By inserting this into the partition function and performing a discrete integration by parts on the lattice, $\sum_{\bm{r}} \theta_{\bm{r}} \nabla \cdot \bm{S}_{\bm{r}} = -\sum_{\ell = \langle \bm{r}_A, \bm{r}_B \rangle} (\theta_{\bm{r}_B} - \theta_{\bm{r}_A}) S^z_\ell$, we find that the \rev{spin sum} factorizes over links:
\begin{equation}
    Z_{v=0} = \int \mathcal{D}\theta \prod_{\langle\bm{r},\bm{r}'\rangle} \left[ 1 + 2w \cos(\theta_{\bm{r}} - \theta_{\bm{r}'}) \right],
\label{eq:angular-zero}
\end{equation}
where $\mathcal{D}\theta \coloneqq \prod_{\bm{r}} \frac{d\theta_{\bm{r}}}{2\pi}$.
\rev{Indeed, expanding each bond factor as $1+w e^{i\nabla\theta_\ell}+w e^{-i\nabla\theta_\ell}$, with $\nabla\theta_\ell=\theta_{\bm r_B}-\theta_{\bm r_A}$, selects an empty link or either current direction. The angle integrals retain only graphs with equal incoming and outgoing currents at every site. Each occupied link carries weight $w$; an isolated loop of length $m$ has two orientations and weight $2w^m$. Thus Eq.~\eqref{eq:angular-zero} is an exact directed-loop representation.}
For sufficiently small $w$, the Taylor expansion $\ln(1+X) \approx X$ holds, yielding
\begin{equation}
    Z_{v=0} \approx \int \mathcal{D}\theta \, e^{2w \sum_{\langle\bm{r},\bm{r}'\rangle} \cos(\theta_{\bm{r}} - \theta_{\bm{r}'})}.
\end{equation}
This is the partition function of the classical 3D $XY$ model defined on the diamond lattice. 
The emergent $XY$ spins interact with an effective ferromagnetic coupling $K = 2w$.
\rev{The leading action motivates a 3D $XY$ description of the continuous transition, consistent with the current arguments and finite-size scaling of Ref.~\cite{Pandey2025}. The omitted harmonics preserve the global U(1) symmetry but modify the bare couplings; the estimate below assumes that the continuous transition remains in the $XY$ basin of attraction.} Using our precise Monte Carlo determination of the critical temperature $T_c = 1.30036(1)$ of the isotropic $XY$ model on the diamond lattice~\cite{Watanabe2026XY}---i.e., the critical coupling $K_c = 1/T_c \approx 0.7690$---we estimate the critical fugacity for the $\mathbb{Z}_2$-deconfined U(1) Coulomb phase onset to be $w_{c_1} = K_c/2 \approx 0.3845$. The deviation from the numerical result $w_{c_1} = 0.3609(1)$ obtained by Pandey and Damle~\cite{Pandey2025} may stem from the higher-order terms in the generalized $XY$ action.

\paragraph{\rev{Unoriented Vacancy Loops and the Ising Model.---}}
\rev{For large $w$, the roles of background and dilute loops are reversed: $S^z=0$ links, of relative weight $1/w$, are unoriented vacancies in an $S^z=\pm1$ ice background}~\cite{Sandvik2006, Jaubert2013}. \rev{The leading small-$w$ action is no longer controlled here. At $v=0$, Gauss's law requires equal numbers of positive and negative links at each site, so the zero-link degree is even (0, 2, or 4). The vacancy graphs are therefore closed, with intersections allowed. These loops lie on diamond-lattice bonds [Fig.~\ref{fig:diamond_pyrochlore}(a)], whose sites host the dual Ising spins. Thus}
\begin{equation}
    Z_{v=0} = w^{2N}W(\emptyset) \sum_{C}\frac{W(C)}{W(\emptyset)}\left(\frac{1}{w}\right)^{|C|},
\end{equation}
where \rev{$|C|$ counts vacancy links and} $W(C)$ is the number of valid background spin configurations for a fixed closed-loop configuration $C$.

\rev{Pauling's approximation treats the ice-rule constraints at different sites independently. Six of the $16$ assignments of four $\pm1$ spins obey the ice rule, giving $W(\emptyset)\approx2^{2N}(3/8)^N$. For a nonintersecting vacancy graph $C$, its $|C|$ zero links traverse $|C|$ sites. The ice-rule probability is $1/2$ there, where the two remaining spins must be opposite, and $3/8$ elsewhere. Hence $W(C)\approx2^{2N-|C|}(3/8)^{N-|C|}(1/2)^{|C|}$ and $W(C)/W(\emptyset)\approx(2/3)^{|C|}$. Extending this line weight to all even-degree graphs while omitting their additional intersection weights gives}
\begin{equation}
    Z_{v=0} \approx w^{2N}W(\emptyset) \sum_{C} \left( \frac{2}{3w} \right)^{|C|}.
\end{equation}

This loop gas is isomorphic to the high-temperature expansion of the zero-field 3D Ising model. In terms of Ising spins $\sigma_{\bm{r}} \in \{-1, 1\}$, the partition function $Z_{\mathrm{Ising}} = \sum_{\{\sigma_{\bm{r}}\}} \exp(K \sum_{\langle \bm{r}, \bm{r}' \rangle} \sigma_{\bm{r}} \sigma_{\bm{r}'})$ can be expanded as
\begin{equation}
    Z_{\mathrm{Ising}} = (\cosh K)^{2N} 2^N \sum_{C} x^{|C|},
\end{equation}
where $x \coloneqq \tanh K$. The proliferation transition of the loop gas, marking the onset of the $\mathbb{Z}_2$-deconfined U(1) Coulomb phase, maps onto the ferromagnetic transition of this Ising model~\cite{Nahum2011}. \rev{The bare weight $x\approx2/(3w)$ relies on Pauling's approximation and neglects background correlations. Intersecting graphs are retained, but their additional factor $(3/2)^{V_4}$ is omitted, where $V_4$ counts degree-four vacancy vertices. The resulting Ising description assumes that these corrections preserve the long-distance fixed point, consistent with the critical behavior observed numerically~\cite{Pandey2025}; its prediction for the critical fugacity is approximate.} \rev{Using the Ising critical coupling} $K_c \approx 0.3698$~\cite{Deng2003} \rev{in} $\frac{2}{3w_{c_2}} \approx \tanh K_c$, we predict $w_{c_2} \approx 1.884$. The agreement with the numerical result $1.8904(1)$~\cite{Pandey2025} demonstrates that this mapping captures the essential physics, with the deviation reflecting the higher-order corrections to Pauling's approximation.

\rev{At $v=0$, dual order diagnoses monopole deconfinement. The operator $e^{iQ\theta_{\bm r}}$ inserts a test charge $Q$; a nonzero long-distance limit of its two-point function implies a finite free-energy cost for separating a test pair. The intermediate phase has dual ferromagnetic order, $\langle e^{i\theta_{\bm r}}\rangle\neq0$, whereas the large-$w$ phase has only nematic order, $\langle e^{2i\theta_{\bm r}}\rangle\neq0$ and $\langle e^{i\theta_{\bm r}}\rangle=0$. These order parameters describe symmetry-broken thermodynamic states in the long-distance dual theory. The source-correlator derivation is given in the Supplemental Material (SM)~\cite{SM}.}

\paragraph{\textcolor{black}{Topological Classification via Global Flux.---}}
At $v=0$, the macroscopic global flux threading the system,
\begin{equation}
    \bm{P} \coloneqq \frac{1}{L}\sum_{\ell = \langle \bm{r}_A, \bm{r}_B \rangle} (\bm{r}_B  - \bm{r}_A )S^z_\ell,
\end{equation}
acts as a topological invariant that distinguishes the three phases. \rev{The bond displacements are nearest-image vectors under periodic boundary conditions.} Since the vertical displacement along any link is $|z_B - z_A| = 1/4$, the system can be partitioned by $4L$ horizontal planes. Let $C_k$ be the set of links intersecting the $k$-th plane. The net flux $W_z^{(k)}$ crossing this plane is given by
\begin{equation}
    W_z^{(k)} = \sum_{\ell \in C_k} \mathrm{sgn}(z_B - z_A) S^z_\ell.
\end{equation}
The ice rule ensures that this flux is conserved across any plane ($W_z^{(k)} = W_z$), yielding $P_z = W_z \in \mathbb{Z}$. By cubic symmetry, $\bm{P} \in \mathbb{Z}^3$.

The parity of $P_z$ can be evaluated modulo $2$. The total number of links crossing a plane is $4L^2$. Since $S^z_\ell \equiv (S^z_\ell)^2 \pmod 2$ and the geometric sign satisfies $\mathrm{sgn}(z_B - z_A) \equiv 1 \pmod 2$, we find $P_z \equiv \sum_{\ell \in C_k} (S^z_\ell)^2 \pmod 2$. If we let $n_0$ be the number of $S^z_\ell = 0$ defect links intersecting the plane, the sum counts the number of nonzero links; thus, the remaining $4L^2 - n_0$ links must have $(S^z_\ell)^2 = 1$. Since $4L^2$ is an even integer, we obtain the congruence relation:
\begin{equation}
    P_z \equiv 4L^2 - n_0 \equiv n_0 \pmod 2.
\end{equation}
\rev{Thus, flux parity equals the parity of the number of zero links crossing a plane. Integer quantization holds exactly at finite $L$, whereas the phase classification concerns sector probabilities in the thermodynamic limit. In the trivial phase ($w<w_{c_1}$), directed $S^z=\pm1$ loops do not proliferate; winding configurations are exponentially suppressed with $L$, leaving $\bm{P}=\bm{0}$. In the $\mathbb{Z}_2$-deconfined U(1) Coulomb phase ($w_{c_1}<w<w_{c_2}$), both directed spin loops and zero-spin strings proliferate, and odd as well as even integer sectors survive. In the $\mathbb{Z}_2$-confined U(1) Coulomb phase ($w>w_{c_2}$), zero-spin strings have tension. Contractible loops cross a plane an even number of times, while winding zero strings and hence odd flux sectors are exponentially suppressed with $L$. The surviving sectors are therefore $(2\mathbb{Z})^3$, with $S^z=\pm1$ loops still proliferating. Thus, the nematic phase confines unit charges while retaining charge-$2$ deconfinement (SM~\cite{SM}).}
This parity-based topological classification parallels the exact description of the 3D toric code limit using macroscopic loop parities~\cite{Nasu2014PRB}. As is evident, this topological nature relies on the absence of monopoles ($v=0$). Once finite-temperature effects are introduced, these topological distinctions become blurred.

\paragraph{\textcolor{black}{Finite-Temperature Effects.---}}
\rev{At low but finite temperature, unit monopoles relax loop closure in both descriptions. For small $w$, $Q=\pm1$ acts as a source or sink of the directed spin currents. In the expansion of the angular weights, a site factor $v e^{\pm i\theta_{\bm{r}}}$ balances the current at such an endpoint. Repeating the mapping to leading order in $w$ and $v$ gives}~\cite{SM}
\begin{equation}
    Z \approx \int \mathcal{D}\theta \, \exp\left(2w \sum_{\langle\bm{r},\bm{r}'\rangle}\cos(\theta_{\bm{r}}-\theta_{\bm{r}'}) + 2v\sum_{\bm{r}}\cos\theta_{\bm{r}}\right).
\label{eq:xy-leading-finite}
\end{equation}
Physically, thermal monopoles introduce a sine-Gordon term acting as a uniform external magnetic field\rev{, $h=2v$ in this leading $XY$ action}. This field explicitly breaks the continuous U(1) symmetry of the dual $XY$ model, rounding the phase transition into a continuous crossover.

\rev{For large $w$, the same unit monopoles create odd-degree vertices in the unoriented vacancy graph, allowing zero-spin strings to end. The partition function is therefore generalized to a sum over graphs $G$ with endpoints,}
\begin{equation}
    Z \approx w^{2N} W(\emptyset) \sum_{G} \left(\frac{2}{3w}\right)^{|G|} (\sqrt{3}v)^{|\partial G|},
\label{eq:graph-finite}
\end{equation}
where $|G|$ is the number of links in the graph, and $|\partial G|$ denotes the total number of endpoints. This matches the high-temperature expansion of the 3D Ising model in a uniform magnetic field $h$, defined as
$Z_{\mathrm{Ising}} = \sum_{\{\sigma_{\bm{r}}\}} \exp(K \sum_{\langle \bm{r}, \bm{r}' \rangle} \sigma_{\bm{r}} \sigma_{\bm{r}'} + h\sum_{\bm{r}} \sigma_{\bm{r}})$.
As detailed in the \rev{SM}~\cite{SM}, this yields
\begin{equation}
    Z_{\mathrm{Ising}} = (\cosh K)^{2N}(\cosh h)^N 2^{N} \sum_{G} x^{|G|} y^{|\partial G|},
\end{equation}
with the effective parameters identified as $x \coloneqq 2/(3w)$ and $y \coloneqq \sqrt{3}v$. 
\rev{Within the Pauling description, $\tanh h=\sqrt3v$ rounds the Ising transition. At either constrained critical point, the corresponding ordering field is relevant: $hL^{y_h}$ grows with system size at fixed $v>0$ because $y_h>0$. Thus, even dilute monopoles cannot be neglected sufficiently close to the monopole-free critical point.}

\rev{The distinction is whether loop closure can be violated. In the 3D toric code, a lattice Bianchi identity forbids magnetic flux endpoints even when thermal electric charges destroy another component of topological order; the closed-loop sector retains a finite-temperature transition~\cite{Castelnovo2008Toric}. Closed flux loops also permit finite-temperature transitions in 3D Kitaev spin liquids~\cite{PhysRevB.90.205126,Nasu2014PRB,Nasu2014}. In spin ice, the divergence constraint is energetic, so thermal monopoles can violate closure.}

\rev{Thermal mobile net charges also screen physical-spin correlations beyond a Debye length $\xi_D$~\cite{Ryzhkin2005,Castelnovo2011}. After the thermodynamic limit, as $v\to0$ at fixed finite noncritical $w$, $\xi_D\propto v^{-1/2}$ in the dual ferromagnet and $v^{-1}$ in the nematic. In the latter, bound same-sign unit charges carry free net charge $2$, with activity $O(v^2/w)$ dominating bare double monopoles ($v^4$). These phase-interior laws are derived in SM Sec.~\ref{sec:screening}. The critical field cutoff instead concerns dual-angle fluctuations near a constrained transition.}

\paragraph{\textcolor{black}{Monte Carlo Simulations.---}}
\begin{figure}[t]
    \centering
    \includegraphics[width=\linewidth]{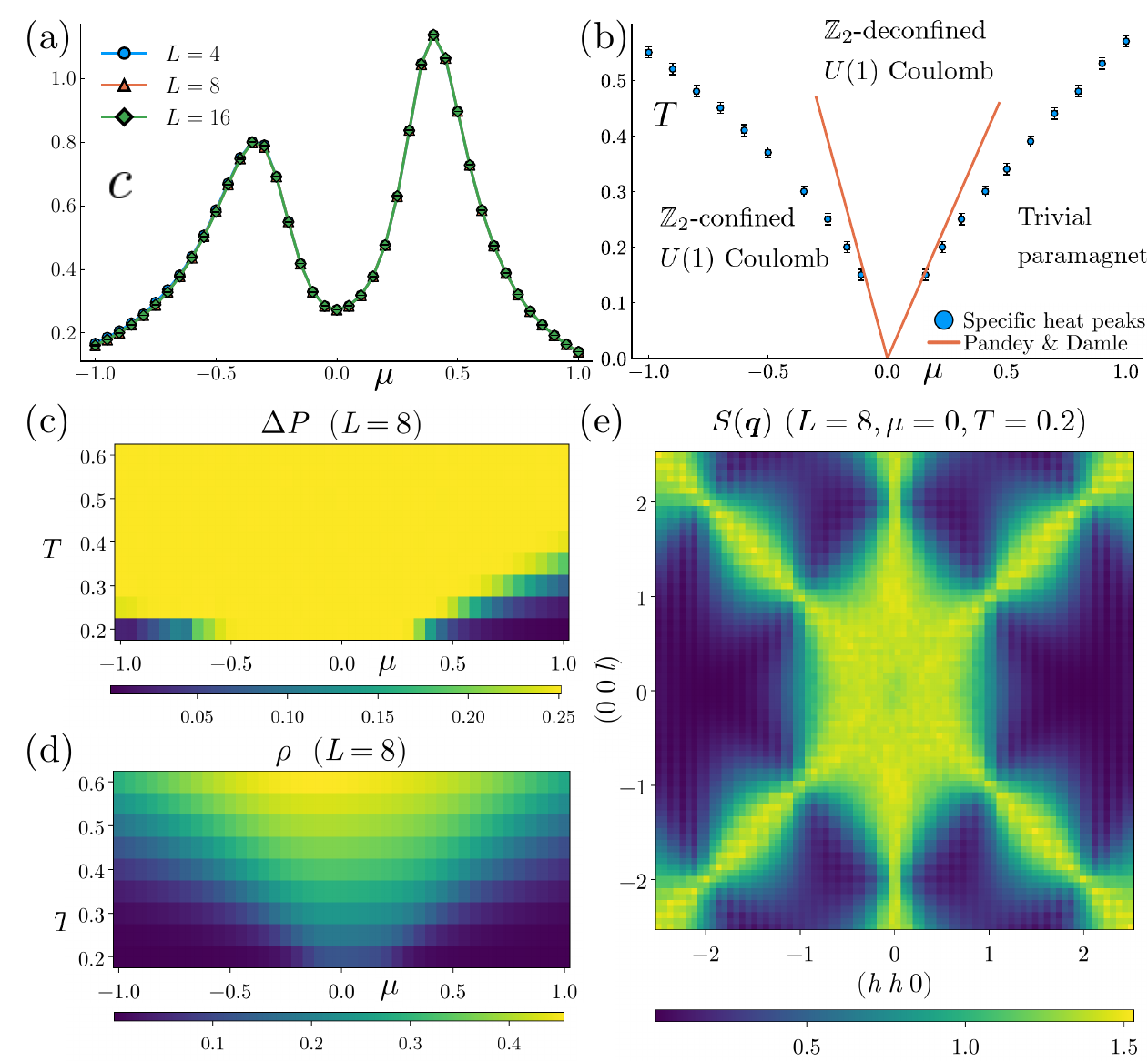}
\caption{Results of MC simulations.
    (a) System-size dependence of the specific heat at $T=0.3$. The three sizes ($L=4,8,16$, distinct symbols) overlap almost perfectly\rev{, showing little size dependence over the range studied [see also Fig.~\ref{fig:magnetic}(b)].}
    (b) \rev{finite-temperature crossover diagram} obtained by peak positions of the specific heat for $L=4$ and the phase boundaries determined numerically in the monopole-free limit by Pandey and Damle~\cite{Pandey2025}.
    Heat maps of (c) the deviation $\Delta P$ and (d) the monopole density $\rho$ for $L=8$. Panels (c) and (d) indicate that the presence of thermal monopoles causes the macroscopic flux to deviate from integer values.
    (e) Spin-flip structure factor~\cite{Fennell2009} $S(\bm{q})$, exhibiting \rev{sharp pinch-point features} characteristic of the $\mathbb{Z}_2$-deconfined U(1) Coulomb regime.
    }
    \label{fig:crossover}
\end{figure}

To verify our predictions, we perform classical MC simulations fixing $J=1$. A standard MC sweep comprises $2N$ single-spin heat-bath updates. Measurements are typically taken over $5 \times 10^6$ sweeps after $10^6$ thermalization sweeps. In the highly frustrated regime (small $\Delta$), we accelerate relaxation using a hybrid scheme combining heat-bath updates with the short-loop algorithm~\cite{Barkema1998, Melko2001} ($10^5$ thermalization, $5 \times 10^5$ measurement sweeps). We calculate the specific heat \rev{per pyrochlore spin},
$c \coloneqq -\frac{T}{\rev{2}N}\frac{\partial^2 F}{\partial T^2} = \frac{1}{\rev{2}N} \frac{\langle E^2 \rangle - \langle E \rangle^2}{T^2}$,
where $E$ is the total energy. We also compute the macroscopic global flux $\bm{P}$ and the monopole density, $\rho\coloneqq\frac{1}{N}\big\langle\sum_{\bm{r}}|\nabla \cdot \bm{S}_{\bm{r}}|\big\rangle$.
We further compute the spin-flip structure factor~\cite{Fennell2009} $S(\bm{q})$ [Fig.~\ref{fig:crossover}(e)]\rev{, the $S^z=0$ occupation $m_0$, and its susceptibility $\chi_0$; their definitions and the $m_0,\chi_0$ data are given in the SM~\cite{SM}. The occupation susceptibility probes the response to the single-ion anisotropy and is distinct from the dual nematic order parameter $\langle e^{2i\theta_{\bm{r}}}\rangle$.}
In the presence of monopoles, the flux $\bm{P}$ is not restricted to $\mathbb{Z}^3$ or $(2\mathbb{Z})^3$.
We define $\Delta P\coloneqq\sum_{i=x,y,z}\big\langle(P_i-\lfloor P_i\rceil)^2\big\rangle$ to measure the deviations of $\bm{P}$ from integer values, where $\lfloor \cdot\rceil$ denotes the nearest integer.

\rev{Figure~\ref{fig:crossover}(a) shows broad heat maxima with little size dependence for $L=4,8,16$. This alone does not establish a crossover, since 3D $XY$ criticality has a finite cusp ($\alpha<0$)~\cite{Campostrini2006}. Finite-$v$ and strict-$v=0$ controls in SM Figs.~\ref{fig:quantitative} and \ref{fig:heat-peak-size}, extending the peak-height comparison to $L=32$, support endpoint-induced rounding. Finite-size scaling of the exact angular weights through $L=80$ is compatible with 3D $XY$ criticality under the stated correction model; finite-$v$ connected dual fluctuations show a cutoff (SM Sec.~\ref{sec:angular-scaling}). The peak positions shift away from the constrained boundaries~\cite{Pandey2025} as $v$ increases [Fig.~\ref{fig:crossover}(b)], defining a crossover diagram.}

\rev{Figures~\ref{fig:crossover}(c,d) show leakage of $\bm P$ from integer quantization accompanied by a finite monopole density, consistent with the loss of the constrained topological distinctions. The sharp pinch-point features in Fig.~\ref{fig:crossover}(e) reflect long screening lengths at the displayed wave-vector resolution. The monopole density alone does not determine $\xi_D$, since neutral bound pairs do not screen net charge.}

\paragraph{\textcolor{black}{Conclusion and Outlook.---}}
\rev{Directed spin loops and unoriented vacancy loops provide complementary $XY$ and Ising descriptions of the constrained $S=1$ pyrochlore ice, with a flux-parity classification summarized in Table~\ref{tab:phases}. Thermal unit monopoles allow both loop types to break and generate relevant dual fields, predicting rounded crossovers near the constrained proliferation transitions. Our MC comparisons support this mechanism in the studied regimes.}

\rev{This framework has also been extended to arbitrary spin $S$~\cite{Watanabe2026SpinS}. Natural extensions include quantum spin ice~\cite{Savary2012,Shannon2012,Benton2012} and quenched dilution~\cite{Sen2015Dilution,Banks2012Disorder}. Unlike such fixed disorder, the present zero-spin occupations fluctuate in equilibrium (annealed dilution). A magnetic realization requires predominantly Ising exchange, a magnetic pair and a nearby nonmagnetic state, $\Delta\simeq J$, and a window $|J_\perp|\ll T\ll J$ with $|\mu|\sim T$, where $J_\perp$ is the transverse exchange~\cite{Pandey2025}. NaCaNi$_2$F$_7$ has nearly Heisenberg spin-$1$ exchange~\cite{Plumb2019,Pohle2025}, while Pr$_2$Zr$_2$O$_7$ and Tb$_2$Ti$_2$O$_7$ have crystal-field doublets~\cite{Kimura2013,Princep2015Tb}; the required three-state Ising regime remains to be established in these platforms. In high-pressure water ice, two off-center and one bond-centered proton positions motivate a related three-state model describing a monopole-rounded ice-VII--ice-X crossover~\cite{Watanabe2026Ice}.}

\begin{acknowledgments}
We thank Naoto Nagaosa, Tibor Rakovszky, and Inti Sodemann for useful discussions.
The work of S.W. is supported by JST SPRING, Grant No.~JPMJSP2108.
The work of Y.M. is supported by JSPS KAKENHI Grant No.~JP25H01247.
The work of H.W. is supported by JSPS KAKENHI Grant No.~JP24K00541.
The work of H.W. was conducted in part during the program ``Generalised symmetries and anomalies in quantum phases of matter 2026'' (code: ICTS/GSYQM2026/01).
\rev{OpenAI Codex assisted with manuscript preparation and the numerical work under the authors' direction.}
\end{acknowledgments}

{\color{black}\section*{Data Availability}}
\rev{The data supporting the findings of this study are available from the authors upon reasonable request.}

\makeatletter
\let\spin@original@lbibitem\@lbibitem
\def\@lbibitem[#1]#2{%
  \bibitem@fin
  \let\@bibstop\@undefined
  \def\spin@bibkey{#2}%
  \color{black}%
  \@for\spin@redbibkey:=Campostrini2006,Sen2015Dilution,Banks2012Disorder,Princep2015Tb,Hasenbusch2019,Powell2012,KentDobias2018,Ballesteros1996,Watanabe2026XY,Watanabe2026Ice,Watanabe2026SpinS\do{%
    \ifx\spin@bibkey\spin@redbibkey\color{black}\fi
  }%
  \spin@original@lbibitem[#1]{#2}%
}
\makeatother
\bibliography{ref_spin1_arXiv3}

\makeatletter\setlength{\@fptop}{0pt}\makeatother
\section*{End Matter}

Table~\ref{tab:phases} collects the three sharply defined ($v=0$) regimes of the $S=1$ pyrochlore spin ice together with their dual $XY$ and Ising characterizations, as established by the mappings of the main text and detailed in the \rev{SM}~\cite{SM}.

\begin{table*}[t]
\caption{Summary of the three sharply defined ($v=0$) regimes of the $S=1$ pyrochlore spin ice and their dual characterizations: the trivial paramagnet ($w<w_{c_1}$), the $\mathbb{Z}_2$-deconfined U(1) Coulomb phase ($w_{c_1}<w<w_{c_2}$), and the $\mathbb{Z}_2$-confined U(1) Coulomb phase ($w>w_{c_2}$). Rows list the dual $XY$ order parameter, the emergent photon, \rev{the flux sectors surviving in the thermodynamic limit (all sectors have integer flux at finite $L$)}, \rev{the large-$w$ dual Ising order, the behavior of directed spin loops and unoriented vacancy loops,} the smallest deconfined (free) monopole charge, \rev{and screening lengths as $v\to0$ at fixed finite noncritical $w$, after the thermodynamic limit. The nematic screening law includes bound same-sign unit-charge pairs. Phase-interior screening is derived in SM Sec.~\ref{sec:screening}; critical field scaling is discussed in Sec.~\ref{sec:angular-scaling}.} \rev{The loop rows describe complementary occupied-spin and zero-spin graphs. Here, confined denotes nonproliferating graphs with positive free-energy line tension. At small $w$, zero-spin links form the background rather than a dilute defect gas; the Pauling-based Ising approximation is outside its regime there.} The onset of the $\mathbb{Z}_2$-deconfined U(1) Coulomb phase at $w_{c_1}$ is a 3D $XY$ transition, and the loop-gas proliferation at $w_{c_2}$ is a 3D Ising transition.}
\label{tab:phases}
{%
\begin{ruledtabular}
\renewcommand{\arraystretch}{1.2}
\begin{tabular}{lccc}
 & $w<w_{c_1}$ & $w_{c_1}<w<w_{c_2}$ & $w>w_{c_2}$ \\
 & trivial paramagnet & $\mathbb{Z}_2$-deconfined U(1) Coulomb & $\mathbb{Z}_2$-confined U(1) Coulomb \\
\colrule
dual $XY$ & disordered & ferromagnet, $\langle e^{i\theta_{\bm{r}}}\rangle\neq0$ & nematic, $\langle e^{2i\theta_{\bm{r}}}\rangle\neq0$, $\langle e^{i\theta_{\bm{r}}}\rangle=0$ \\
emergent photon & none (gapped) & gapless & gapless \\
\rev{surviving flux sectors} & $\bm{0}$ & $\mathbb{Z}^3$ & $(2\mathbb{Z})^3$ \\
dual Ising & \rev{outside its regime} & ordered, $\langle\sigma_{\bm{r}}\rangle\neq0$ & disordered \\
\rev{directed $S^z=\pm1$ loops} & \rev{confined} & \rev{proliferate} & \rev{proliferate} \\
\rev{unoriented $S^z=0$ loops} & \rev{background} & proliferate & \rev{confined} \\
smallest free charge & --- & $Q=1$ & $Q=2$ \\
\rev{$\xi_D$ (fixed finite $w$)} & --- & $v^{-1/2}$ & \rev{$v^{-1}$} \\
\colrule
transition & \multicolumn{1}{r}{\rlap{\hspace{\tabcolsep}\makebox[0pt][c]{$w_{c_1}$: 3D $XY$}}} & \multicolumn{1}{r}{\rlap{\hspace{\tabcolsep}\makebox[0pt][c]{$w_{c_2}$: 3D Ising}}} & \\
\end{tabular}
\end{ruledtabular}
}
\end{table*}

\clearpage
\setcounter{section}{0}
\onecolumngrid

\clearpage
\section*{SUPPLEMENTAL MATERIAL}
\setcounter{equation}{0}
\renewcommand{\theequation}{S\arabic{equation}}
\renewcommand{\theHequation}{S\arabic{equation}}
\setcounter{figure}{0}
\renewcommand{\thefigure}{S\arabic{figure}}
\renewcommand{\theHfigure}{S\arabic{figure}}
\setcounter{table}{0}
\renewcommand{\thetable}{S\arabic{table}}
\renewcommand{\theHtable}{S\arabic{table}}
\makeatletter
\renewcommand{\fnum@figure}{\ifnum\value{figure}>1\relax\textcolor{black}{FIG.~\thefigure}\else\textcolor{black}{FIG.~\thefigure}\fi}
\renewcommand{\fnum@table}{\textcolor{black}{TABLE~\thetable}}
\makeatother
\section{Order-Parameter Structure of the $\mathbb{Z}_2$-Confined U(1) Coulomb Phase}

\rev{We work at $v=0$ and derive charge-$1$ and charge-$2$ source correlations that distinguish the dual ferromagnetic and nematic orders. The exact angular representation preserves global U(1); its weight is signed for $w>1/2$, so the order statements refer to the infrared dual theory and source-insertion correlations of the original positive spin ensemble. In symmetry-selected thermodynamic states, the ordered regimes are}
\begin{equation}
\text{ferromagnet:}\ \langle e^{i\theta_{\bm{r}}}\rangle\neq0\ (\text{rank-1, polar}),
\qquad
\text{nematic:}\ \langle e^{i\theta_{\bm{r}}}\rangle=0,\ \langle e^{2i\theta_{\bm{r}}}\rangle\neq0\ (\text{rank-2, director}),
\end{equation}
in analogy with a liquid crystal ($e^{i\theta_{\bm{r}}}$ is a headed vector, $e^{2i\theta_{\bm{r}}}$ a headless director). Both break the global $U(1)$---the nematic only down to $\mathbb{Z}_2$ ($\theta_{\bm{r}}\to\theta_{\bm{r}}+\pi$)---so both retain a gapless Goldstone mode, the emergent photon, and are Coulomb liquids with pinch-point singularities. \rev{The $\mathbb{Z}_2$-deconfined U(1) Coulomb phase ($w_{c_1}<w<w_{c_2}$) is the ferromagnet (odd and even flux sectors survive); the $\mathbb{Z}_2$-confined U(1) Coulomb phase ($w>w_{c_2}$) is the nematic (only even sectors survive thermodynamically)~\cite{Pandey2025}. In the nematic regime, odd sectors have exponentially small, nonzero weight at finite $L$ and finite $w$.} The transition at $w_{c_2}$ is therefore the ferromagnet-to-nematic transition of the \rev{dual angular theory}---the restoration of the $\mathbb{Z}_2$ symmetry $\theta_{\bm{r}}\to\theta_{\bm{r}}+\pi$---\rev{whose Ising description is consistent with the loop-gas mapping, provided corrections to Pauling's approximation preserve its long-distance fixed point.} \rev{At strictly $v=0$, the surviving charge-$2$ (rank-$2$) order belongs to the dual phase field and characterizes an emergent topological distinction in the original spin ensemble. It is not a local magnetic quadrupolar order parameter.}

\rev{At $v>0$, induced polar order and thermal string breaking preclude strict charge-$1$ confinement at arbitrarily large distance; thermal screening is discussed in Sec.~\ref{sec:screening}.}

\subsection{Test-charge correlator: an exact identity}
{%
Because no dynamical monopoles exist at $v=0$, every $S^z=0$ string is closed, and the confinement of the charge-$1$ excitation is probed by an \emph{external} static test pair rather than by a thermally excited string. For a fixed background charge pattern $\rho=\{\rho_{\bm{r}}\}$, we insert it as a source into the $v=0$ partition function and define the constrained partition function
\begin{equation}
Z(\rho)\coloneqq\sum_{\{S^z_\ell\}}\Big(\prod_{\bm{r}}\delta_{\nabla\cdot\bm{S}_{\bm{r}},\,\rho_{\bm{r}}}\Big)\,w^{\sum_\ell(S^z_\ell)^2},
\end{equation}
so that $Z(0)=Z_{v=0}$ recovers the physical partition function. \rev{We first derive its exact angular representation, then estimate the loop-gas weights using Pauling's approximation to relate the source correlator to the Ising model.}

We first evaluate $Z(\rho)$ \rev{using the angular representation}. Representing each Gauss-law constraint by $\delta_{\nabla\cdot\bm{S}_{\bm{r}},\rho_{\bm{r}}}=\int_0^{2\pi}\frac{d\theta_{\bm{r}}}{2\pi}\,e^{-i\theta_{\bm{r}}\nabla\cdot\bm{S}_{\bm{r}}+i\theta_{\bm{r}}\rho_{\bm{r}}}$ and integrating by parts, $\sum_{\bm{r}}\theta_{\bm{r}}\nabla\cdot\bm{S}_{\bm{r}}=-\sum_\ell\nabla\theta_\ell S^z_\ell$, the source enters only through a pure site factor while the link sum factorizes independently of it. With the exact dual link weight
\begin{equation}
\mathcal{W}[\theta]\coloneqq\prod_\ell\big(1+2w\cos\nabla\theta_\ell\big),\qquad \nabla\theta_\ell=\theta_{\bm{r}_B}-\theta_{\bm{r}_A},
\end{equation}
the constrained partition function reads
\begin{equation}
Z(\rho)=\int\mathcal{D}\theta\Big(\prod_{\bm{r}}e^{i\theta_{\bm{r}}\rho_{\bm{r}}}\Big)\prod_\ell\sum_{S^z_\ell\in\{0,\pm1\}}e^{iS^z_\ell\nabla\theta_\ell}w^{(S^z_\ell)^2}=\int\mathcal{D}\theta\Big(\prod_{\bm{r}}e^{i\theta_{\bm{r}}\rho_{\bm{r}}}\Big)\mathcal{W}[\theta],
\end{equation}
where $\mathcal{D}\theta=\prod_{\bm{r}}d\theta_{\bm{r}}/2\pi$. Normalizing by $Z(0)$ defines the dual average $\langle\mathcal{O}\rangle\coloneqq\int\mathcal{D}\theta\,\mathcal{O}\,\mathcal{W}[\theta]\big/\!\int\mathcal{D}\theta\,\mathcal{W}[\theta]$. For the unit test source $\rho^{(12)}_{\bm{r}}\coloneqq\delta_{\bm{r},\bm{r}_1}-\delta_{\bm{r},\bm{r}_2}$ [a $+1$ test charge at $\bm{r}_1$ and a $-1$ at $\bm{r}_2$], the site factor collapses to $\prod_{\bm{r}}e^{i\theta_{\bm{r}}\rho^{(12)}_{\bm{r}}}=e^{i\theta_{\bm{r}_1}}e^{-i\theta_{\bm{r}_2}}$, so that
\begin{equation}
\langle e^{i\theta_{\bm{r}_1}}e^{-i\theta_{\bm{r}_2}}\rangle=\frac{Z(\rho^{(12)})}{Z(0)}
\label{eq:xy-route}
\end{equation}
holds \emph{exactly} for all $w$. This is an algebraic identity that does not require $\mathcal{W}[\theta]$ to be positive (it is not, for $w>1/2$), since $Z(\rho^{(12)}),Z(0)>0$.

We next evaluate the same $Z(\rho)$ within the dual loop-gas mapping to the Ising model. In those variables [defect graph $G=\{\ell:S^z_\ell=0\}$, degree $d_0(\bm{r})$, boundary $\partial G=\{\bm{r}:d_0(\bm{r})\ \mathrm{odd}\}$], a configuration with source $\rho$ carries weight $w^{2N-|G|}$ and a multiplicity $W(\rho,G)$ equal to the purely combinatorial (temperature-independent) number of $\pm1$ backgrounds on the links $\ell\notin G$ that realize the charge pattern $\rho$,
\begin{equation}
W(\rho,G)\coloneqq\sum_{\{S^z_\ell\}}\Big(\prod_{\bm{r}}\delta_{\nabla\cdot\bm{S}_{\bm{r}},\,\rho_{\bm{r}}}\Big)\Big(\prod_{\ell\in G}\delta_{S^z_\ell,0}\Big)\Big(\prod_{\ell\notin G}\delta_{|S^z_\ell|,1}\Big).
\label{eq:Wrhodef}
\end{equation}
Because the parity of $\nabla\cdot\bm{S}_{\bm{r}}$ equals that of $d_0(\bm{r})$, $W(\rho,G)$ vanishes unless the defect endpoints coincide with the sources, so only graphs with $\partial G=\partial\rho$ contribute:
\begin{equation}
Z(\rho)=\sum_{G:\,\partial G=\partial\rho}w^{2N-|G|}\,W(\rho,G),\qquad \partial\rho\coloneqq\{\bm{r}:\rho_{\bm{r}}\ \mathrm{odd}\}.
\label{eq:Zrho-loop}
\end{equation}
For the test pair $\rho^{(12)}$ the numerator runs over open $S^z=0$ strings with endpoints $\{\bm{r}_1,\bm{r}_2\}$ \rev{(each endpoint has an odd number of nonzero spin links)}, and the common prefactor $w^{2N}$ cancels between numerator and denominator. \rev{Within the local Pauling approximation, a fixed signed unit charge has three allowed configurations at a degree-1 endpoint and one at a degree-3 endpoint, half the counts summed over the two signs for a thermal monopole. With $V_k$ the number of degree-$k$ vertices,
\[
\frac{W(\rho,G)}{W(\emptyset)}\approx
\left(\frac23\right)^{|G|}
\left(\frac{\sqrt3}{2}\right)^{|\partial G|}
\left(\frac32\right)^{V_4}.
\]
For distinct unit test charges, $|\partial G|=2$. Dropping the self-intersection factor and writing $x\coloneqq\tanh K=2/(3w)$ gives}
\begin{equation}
\underbrace{\frac{Z(\rho^{(12)})}{Z(0)}=\frac{\sum_{\partial G=\{\bm{r}_1,\bm{r}_2\}}w^{-|G|}W(\rho^{(12)},G)}{\sum_{\partial G=\emptyset}w^{-|G|}W(0,G)}}_{\text{exact}}\;\underbrace{\approx\;\rev{\frac34}\frac{\sum_{\partial G=\{\bm{r}_1,\bm{r}_2\}}x^{|G|}}{\sum_{\partial G=\emptyset}x^{|G|}}=\rev{\frac34}\langle\sigma_{\bm{r}_1}\sigma_{\bm{r}_2}\rangle_{\mathrm{Ising}}}_{\text{Pauling}},
\label{eq:loop-route}
\end{equation}
where
\begin{equation}
\langle\sigma_{\bm{r}_1}\sigma_{\bm{r}_2}\rangle_{\mathrm{Ising}}\coloneqq\frac{1}{Z_{\mathrm{Ising}}}\sum_{\{\sigma_{\bm{r}}\}}\sigma_{\bm{r}_1}\sigma_{\bm{r}_2}\,e^{K\sum_{\langle\bm{r},\bm{r}'\rangle}\sigma_{\bm{r}}\sigma_{\bm{r}'}},\qquad Z_{\mathrm{Ising}}=\sum_{\{\sigma_{\bm{r}}\}}e^{K\sum_{\langle\bm{r},\bm{r}'\rangle}\sigma_{\bm{r}}\sigma_{\bm{r}'}},
\end{equation}
is the two-point function of the zero-field 3D Ising model of the main text ($\sigma_{\bm{r}}=\pm1$ on the diamond sites, $x=\tanh K$). The last equality is its high-temperature expansion: with $e^{K\sigma_{\bm{r}}\sigma_{\bm{r}'}}=\cosh K\,(1+x\,\sigma_{\bm{r}}\sigma_{\bm{r}'})$, expanding the product over the $2N$ bonds labels each term by the subgraph $G$ of bonds carrying the factor $x\,\sigma_{\bm{r}}\sigma_{\bm{r}'}$, with weight $x^{|G|}\prod_{\bm{r}}\sigma_{\bm{r}}^{d_0(\bm{r})}$ [$d_0(\bm{r})$ the degree of $\bm{r}$ in $G$]. Since $\sum_{\sigma_{\bm{r}}=\pm1}\sigma_{\bm{r}}^{m}=2$ for even $m$ and $0$ for odd $m$, the spin sum annihilates every graph with an odd-degree site; thus $Z_{\mathrm{Ising}}=(\cosh K)^{2N}2^N\sum_{\partial G=\emptyset}x^{|G|}$ keeps only closed graphs, while the extra $\sigma_{\bm{r}_1}\sigma_{\bm{r}_2}$ in the numerator flips the required parity at $\bm{r}_1,\bm{r}_2$ and keeps instead the graphs with $\partial G=\{\bm{r}_1,\bm{r}_2\}$. The common prefactor $(\cosh K)^{2N}2^N$ cancels in the ratio, establishing the last equality of Eq.~\eqref{eq:loop-route}.

\rev{Combining the exact angular identity, Eq.~\eqref{eq:xy-route}, with the approximate loop-gas relation, Eq.~\eqref{eq:loop-route}, yields the Ising correlator with an endpoint amplitude,}
\begin{equation}
\langle e^{i\theta_{\bm{r}_1}}e^{-i\theta_{\bm{r}_2}}\rangle\mathrel{\rev{\approx}}\rev{\frac34}\langle\sigma_{\bm{r}_1}\sigma_{\bm{r}_2}\rangle_{\mathrm{Ising}}.
\label{eq:xy-ising}
\end{equation}
\rev{The constant endpoint amplitude does not affect the long-distance confinement criterion. Background correlations and self-intersections can renormalize this amplitude and the critical fugacity; the Ising universality assignment assumes that these corrections preserve the long-distance fixed point.}

This common ratio is the single-monopole propagator; writing it as
\begin{equation}
\langle e^{i\theta_{\bm{r}_1}}e^{-i\theta_{\bm{r}_2}}\rangle=\frac{Z(\rho^{(12)})}{Z(0)}=e^{-\Delta F(\bm{r}_1,\bm{r}_2)/T}
\label{eq:F-def}
\end{equation}
identifies $\Delta F(\bm{r}_1,\bm{r}_2)$ as the free-energy cost of separating the test pair. In the $\mathbb{Z}_2$-deconfined U(1) Coulomb phase ($w_{c_1}<w<w_{c_2}$) the propagator saturates, $\langle e^{i\theta_{\bm{r}_1}}e^{-i\theta_{\bm{r}_2}}\rangle\to\mathrm{const}$, so the charge-$1$ monopole is \emph{deconfined}; \rev{its nonzero insertion amplitude gives a free-carrier density of order $v$, and hence $\xi_D\propto v^{-1/2}$ as $v\to0$ at fixed noncritical $w$, after the thermodynamic limit (Sec.~\ref{sec:screening}).} In the nematic phase ($w>w_{c_2}$) the propagator decays, $\langle e^{i\theta_{\bm{r}_1}}e^{-i\theta_{\bm{r}_2}}\rangle\sim e^{-|\bm{r}_1-\bm{r}_2|/\xi_\parallel}$ with a confinement length $\xi_\parallel$ set by the $S^z=0$ string tension, so the charge-$1$ monopole is \emph{confined}---matching the charge-$1$ worm correlator of Pandey and Damle~\cite{Pandey2025}. \rev{This confinement statement is at $v=0$; the odd-charge confinement length is distinct from the finite-$v$ Debye screening length.} \rev{The decay of this physical odd-charge source correlator and the thermodynamic suppression of odd winding sectors are complementary diagnostics of the transition described by the Ising mapping.}
}

\subsection{The complementary nematic order parameter}
{%
More generally, the same manipulation gives the charge-$Q$ propagator
\begin{equation}
\langle e^{iQ\theta_{\bm{r}_1}}e^{-iQ\theta_{\bm{r}_2}}\rangle=\frac{Z(Q\rho^{(12)})}{Z(0)}=e^{-\Delta F_Q(\bm{r}_1,\bm{r}_2)/T},
\label{eq:GQ-def}
\end{equation}
with $\Delta F_Q(\bm{r}_1,\bm{r}_2)$ the free-energy cost of the charge-$Q$ test pair [Eq.~\eqref{eq:F-def} is the case $Q=1$, $\Delta F_1(\bm{r}_1,\bm{r}_2)=\Delta F(\bm{r}_1,\bm{r}_2)$]. Accordingly the charge-$Q$ monopole is \emph{free} (deconfined) when $\lim_{|\bm{r}_1-\bm{r}_2|\to\infty}\langle e^{iQ\theta_{\bm{r}_1}}e^{-iQ\theta_{\bm{r}_2}}\rangle>0$ and \emph{confined} when it decays as $e^{-\sigma_Q|\bm{r}_1-\bm{r}_2|}$ ($\sigma_Q>0$). \rev{The nematic order parameter lives in the charge-$2$ channel. We therefore examine how its source correlator differs from the charge-$1$ correlator.}

The loop-gas form Eq.~\eqref{eq:Zrho-loop} applies to $\langle e^{iQ\theta_{\bm{r}_1}}e^{-iQ\theta_{\bm{r}_2}}\rangle$ as well, and a parity argument sharply separates odd from even $Q$. In a pure $S^z=\pm1$ background the site charge $\nabla\cdot\bm{S}_{\bm{r}}=\sum_{\ell\in\bm{r}}S^z_\ell$ is a sum of four odd numbers and is thus always \emph{even}; an odd charge can arise only where an $S^z=0$ link leaves an odd number of $\pm1$ links, i.e., at the endpoint of an open $S^z=0$ string. Hence $\partial(Q\rho^{(12)})=\{\bm{r}_1,\bm{r}_2\}$ for odd $Q$---the numerator then requires an $S^z=0$ string joining the two sources---while $\partial(Q\rho^{(12)})=\emptyset$ for even $Q$. The charge-$2$ propagator is accordingly a ratio over \emph{closed} defect graphs,
\begin{equation}
\langle e^{2i\theta_{\bm{r}_1}}e^{-2i\theta_{\bm{r}_2}}\rangle=\frac{Z(2\rho^{(12)})}{Z(0)}=\frac{\sum_{\partial G=\emptyset}w^{-|G|}\,W(2\rho^{(12)},G)}{\sum_{\partial G=\emptyset}w^{-|G|}\,W(0,G)},
\label{eq:loop-route-Q2}
\end{equation}
\rev{Equation~\eqref{eq:loop-route-Q2} is an exact decomposition over closed defect graphs. Unlike the charge-$1$ numerator of Eq.~\eqref{eq:loop-route}, it does not force an open zero-spin string between the sources. It is not an exact reduction to the correlator of undiluted $S^z=\pm1$ ice: each graph changes the allowed background configurations and their correlations.}

\rev{At $w=\infty$, the sources are the usual $Q=\pm2$ monopoles of $S^z=\pm1$ pyrochlore ice, whose separation has a finite asymptotic free-energy cost in its Coulomb phase. Extending this deconfinement to finite $w$ assumes stability of that Coulomb phase under closed zero-spin loops. Deep in the large-$w$ phase the loops are dilute, but that description need not hold close to $w_{c_2}$. The persistence of charge-$2$ deconfinement throughout the nematic regime is supported by the charge-$2$ worm correlations of Pandey and Damle~\cite{Pandey2025}, although their worm histogram is not a direct estimator of the source ratio in Eq.~\eqref{eq:loop-route-Q2}. Thus the parity construction explains why an odd-charge string tension does not imply confinement of even charges; the long-distance conclusion additionally uses Coulomb-phase stability and the numerical evidence.}

\rev{In the dual ferromagnet, charge-$1$ order also permits charge-$2$ order, whereas both channels are short-ranged below $w_{c_1}$. The smallest deconfined net charge in the nematic reference state is therefore $Q=2$.}
}

\section{Dual Mapping to the $XY$ Model at Finite Temperatures}
In the main text, we introduced the duality mapping to the classical $XY$ model in the monopole-free limit ($v=0$). Here, we generalize this mapping to finite temperatures, where the local constraints are relaxed and thermal monopoles with local charge $Q_{\bm{r}} = \nabla \cdot \bm{S}_{\bm{r}} \ne 0$ are thermally excited.

To enforce this generalized local constraint, we introduce a continuous phase field $\theta_{\bm{r}} \in [0, 2\pi)$ at each diamond-lattice site $\bm{r}$ via the Fourier integral representation of the Kronecker delta:
\begin{equation}
    \delta_{\nabla \cdot \bm{S}_{\bm{r}}, Q_{\bm{r}}} = \int_0^{2\pi} \frac{d\theta_{\bm{r}}}{2\pi} e^{-i \theta_{\bm{r}} \nabla \cdot \bm{S}_{\bm{r}} + i \theta_{\bm{r}} Q_{\bm{r}}}.
\end{equation}
By inserting this integral into the partition function $Z$ and performing a discrete integration by parts on the lattice, $\sum_{\bm{r}} \theta_{\bm{r}} \nabla \cdot \bm{S}_{\bm{r}} = -\sum_{\ell} \nabla \theta_{\ell} S^z_\ell$ (where $\nabla \theta_{\ell} \coloneqq \theta_{\bm{r}_B} - \theta_{\bm{r}_A}$), \rev{we can perform the spin and charge sums independently at fixed $\theta$. The resulting integrand is a product of link and site factors:}
\begin{align}
    Z &= \sum_{\{Q_{\bm{r}}\}}\sum_{\{S_\ell^z\}} \prod_{\bm{r}} \left( \int_0^{2\pi} \frac{d\theta_{\bm{r}}}{2\pi} e^{-i \theta_{\bm{r}} \nabla \cdot \bm{S}_{\bm{r}} + i \theta_{\bm{r}} Q_{\bm{r}}} \right) v^{\sum_{\bm{r}} Q_{\bm{r}}^2} w^{\sum_\ell (S_\ell^z)^2} \notag \\
    &= \int \mathcal{D}\theta \prod_{\ell} \left( \sum_{S_\ell^z \in \{0, \pm 1\}} e^{i S_\ell^z \nabla \theta_{\ell}} w^{(S_\ell^z)^2} \right) \prod_{\bm{r}} \left( \sum_{Q_{\bm{r}}} v^{Q_{\bm{r}}^2} e^{i Q_{\bm{r}} \theta_{\bm{r}}} \right) \notag \\
    &= \int \mathcal{D}\theta \prod_{\langle\bm{r},\bm{r}'\rangle} \left[ 1 + 2w \cos(\theta_{\bm{r}} - \theta_{\bm{r}'}) \right] \prod_{\bm{r}} \left[ 1 + 2\sum_{Q_{\bm{r}} > 0} v^{Q_{\bm{r}}^2} \cos(Q_{\bm{r}} \theta_{\bm{r}}) \right],
\label{eq:angular-finite}
\end{align}
where $\mathcal{D}\theta \coloneqq \prod_{\bm{r}} \frac{d\theta_{\bm{r}}}{2\pi}$ is the integration measure. Note that $Q_{\bm{r}}$ takes integer values ranging from $-4$ to $4$, representing the discrete divergence of four spins at a site.

\rev{The loop content follows directly by expanding each link weight as $1+w e^{i\nabla\theta_\ell}+w e^{-i\nabla\theta_\ell}$. Let $\mathcal C$ denote the resulting oriented occupied-link graph, with current $j_\ell=0,\pm1$, number of occupied links $|\mathcal C|=\sum_\ell j_\ell^2$, and divergence $q_{\bm r}(\mathcal C)=(\nabla\cdot\bm j)_{\bm r}$. The site Fourier sum and angle integral enforce $Q_{\bm r}=q_{\bm r}$, giving the exact graph weights}
{\color{black}
\[
Z_{v=0}=\sum_{\mathcal C:\,q_{\bm r}=0\ \forall\bm r}w^{|\mathcal C|},\qquad
Z=\sum_{\mathcal C}w^{|\mathcal C|}v^{\sum_{\bm r}q_{\bm r}(\mathcal C)^2}.
\]
}
\rev{Each current configuration is counted once. At $v=0$, it decomposes into closed directed cycles, with nonunique routing at intersections. Nonintersecting loops contribute $2^{n_{\rm loop}}w^{|\mathcal C|}$ after summing orientations. At $v>0$, nonzero divergence marks sources and sinks; an open unit-current path of length $m$ has weight $w^m v^2$ per orientation.}

\rev{For $w\ll1$ and $v\ll1$, expanding the logarithms of these exact weights to leading order and retaining the unit-charge term gives}
\begin{equation}
    Z \approx \int \mathcal{D}\theta \, \exp\left[ 2w \sum_{\langle\bm{r},\bm{r}'\rangle} \cos(\theta_{\bm{r}} - \theta_{\bm{r}'}) + 2v \sum_{\bm{r}} \cos\theta_{\bm{r}} \right].
\end{equation}
\rev{Expanding this leading action yields the same current-conservation and endpoint rules, although the graph weights differ from the exact restricted-current sum above. The monopole term is a uniform external magnetic field (a sine-Gordon potential) coupled to the emergent $XY$ phase variables: the geometric allowance of endpoints and explicit breaking of the dual symmetry are the same perturbation.}

{\color{black}\subsection{Screening at small monopole fugacity}\label{sec:screening}}

\rev{The exact weights also determine dilute-carrier screening beyond the small-$w$ approximation. Here the spin-screening length $\xi_D$ is the inverse screening wave number, or mass length of the dual phase mode. For dilute mobile carriers of net charge $Q_a$ and density $n_a$, $\xi_D^{-2}\propto\sum_a Q_a^2n_a$, with background stiffness absorbed in the coefficient~\cite{Castelnovo2011}. Neutral bound pairs do not contribute to this sum. Taking the thermodynamic limit first and then $v\to0$ at fixed finite $w$ away from either transition, the ferromagnetic reference state has free unit-charge carriers of density $O(v)$, giving $\xi_D\propto v^{-1/2}$.}

\rev{In the nematic reference state, unit charges are confined but net charge $2$ is deconfined. A mobile charged composite is obtained from adjacent charges $(+2,0)$ by setting their shared outward-unit-flux link to $S^z=0$, producing $(+1,+1)$. The net charge is preserved, while its monopole energy decreases from $2J$ to $J$ and removing the occupied link costs $-\mu$. Its activity relative to the ice background is thus $e^{-(J-\mu)/T}=v^2/w$. The short same-sign pair can separate from a remote negative carrier without extending its binding string. At finite $w$, these $O(v^2)$ carriers precede bare double monopoles of activity $v^4$, giving $\xi_D\propto v^{-1}$. These phase-interior screening laws differ from the critical field cutoff in Sec.~\ref{sec:angular-scaling}; the near-$XY$ simulations do not measure nematic screening.}

\section{Mapping to the Ising Model with a Magnetic Field}
To investigate the effects of thermal monopoles ($v>0$) in the large-$w$ regime, we generalize Pauling's approximation to configurations with nonzero monopole charges and establish a term-by-term correspondence with the high-temperature expansion of the 3D Ising model in a magnetic field.

\rev{This representation is complementary to the occupied-current graphs of Sec.~II: the background is now $S^z=\pm1$ ice and the graph records unoriented $S^z=0$ links. Since $Q_{\bm r}\equiv d_0(\bm r)\pmod2$, odd-charge monopoles allow odd-degree endpoints, while even charges modify background weights without breaking this parity closure. Unit monopoles therefore permit endpoints in both descriptions, with orientation retained only in the occupied-current representation.} The partition function can be expressed as a sum over general graph configurations $G$ that may contain endpoints and self-intersections:
\begin{equation}
    Z = w^{2N} \sum_{G} \left(\frac{1}{w}\right)^{|G|} W_v(G),
\end{equation}
where $W_v(G)$ is the weighted number of valid background spin configurations for a fixed graph $G$, defined by
\begin{equation}
    W_v(G) \coloneqq \sum_{\{Q_{\bm{r}}\}} \sum_{\{S_\ell^z\}} \left( \prod_{\bm{r}} \delta_{\nabla \cdot \bm{S}_{\bm{r}}, Q_{\bm{r}}} \right) \left( \prod_{\ell \in G} \delta_{S_\ell^z, 0} \right) \left( \prod_{\ell \notin G} \delta_{|S_\ell^z|, 1} \right) v^{\sum_{\bm{r}} Q_{\bm{r}}^2}.
\end{equation}

Let $d_0(\bm{r})$ be the number of defect links (fixed to $S_\ell^z = 0$) connected to a given site $\bm{r}$. The remaining $4 - d_0(\bm{r})$ links can freely take $\pm 1$. We define the local state sum $z(d_0)$ by summing the statistical weight $v^{Q^2}$ over all $2^{4 - d_0}$ possible assignments:
\begin{equation}
    z(d_0) \coloneqq \sum_{Q} \binom{4 - d_0}{\frac{4 - d_0 + Q}{2}} v^{Q^2}.
\end{equation}
In the low-temperature regime ($v \ll 1$), we neglect higher-order monopole contributions (e.g., $v^4$). The local state sums are evaluated as:
\begin{align}
    &d_0=0: \quad Q=0 \text{ dominates, giving } z(0) \approx \binom{4}{2} v^0 = 6. \notag \\
    &d_0=1: \quad Q=\pm 1, \text{ giving } z(1) \approx 2\binom{3}{1} v = 6v. \notag \\
    &d_0=2: \quad Q=0 \text{ dominates, giving } z(2) \approx \binom{2}{1} v^0 = 2. \notag \\
    &d_0=3: \quad Q=\pm 1, \text{ giving } z(3) \approx 2\binom{1}{0} v = 2v. \notag \\
    &d_0=4: \quad Q=0, \text{ giving } z(4) = \binom{0}{0} v^0 = 1.
\end{align}

Under Pauling's approximation, the total weight $W_v(G)$ is estimated by multiplying the $2^{2N - |G|}$ completely free degrees of freedom by the probabilities $z(d_0) / 2^{4 - d_0}$ for realizing the specific charge configurations at all sites:
\begin{equation}
    W_v(G) \approx 2^{2N - |G|} \prod_{\bm{r}} \frac{z(d_0(\bm{r}))}{2^{4 - d_0(\bm{r})}} = 2^{|G| - 2N} \prod_{\bm{r}} z(d_0(\bm{r})),
\end{equation}
where we used the identity $\sum_{\bm{r}} [4 - d_0(\bm{r})] = 4N - 2|G|$. Let $V_k$ denote the number of sites in graph $G$ with degree $k$ (i.e., $d_0(\bm{r}) = k$). The product over sites becomes $z(0)^{V_0} z(1)^{V_1} z(2)^{V_2} z(3)^{V_3} z(4)^{V_4}$. Factoring out the vacuum weight $W(\emptyset) \approx 2^{-2N} z(0)^N$ and using $V_0 = N - V_1 - V_2 - V_3 - V_4$, we obtain
\begin{align}
    W_v(G) &\approx W(\emptyset) 2^{|G|} \left(\frac{z(1)}{z(0)}\right)^{V_1} \left(\frac{z(2)}{z(0)}\right)^{V_2} \left(\frac{z(3)}{z(0)}\right)^{V_3} \left(\frac{z(4)}{z(0)}\right)^{V_4} \notag \\
    &= W(\emptyset) 2^{|G|} (v)^{V_1} \left(\frac{1}{3}\right)^{V_2} \left(\frac{v}{3}\right)^{V_3} \left(\frac{1}{6}\right)^{V_4}.
\end{align}
We eliminate $V_2$ by exploiting the graph-theoretic identity for the sum of degrees: $V_1 + 2V_2 + 3V_3 + 4V_4 = 2|G| \implies V_2 = |G| - \frac{1}{2}V_1 - \frac{3}{2}V_3 - 2V_4$. This yields
\begin{align}
    W_v(G) &\approx W(\emptyset) 2^{|G|} v^{V_1} \left[ \left(\frac{1}{3}\right)^{|G|} (\sqrt{3})^{V_1} (3\sqrt{3})^{V_3} 9^{V_4} \right] \left(\frac{v}{3}\right)^{V_3} \left(\frac{1}{6}\right)^{V_4} \notag \\
    &= W(\emptyset) \left(\frac{2}{3}\right)^{|G|} (\sqrt{3}v)^{V_1 + V_3} \left(\frac{3}{2}\right)^{V_4}.
\end{align}
The sum of odd-degree vertices $V_1 + V_3$ corresponds to the total number of endpoints $|\partial G|$ in the graph. Assuming a dilute gas where self-intersections ($V_4$) are rare, we approximate $(3/2)^{V_4} \approx 1$. The partition function then simplifies to
\begin{equation}
    Z \approx w^{2N} W(\emptyset) \sum_{G} \left(\frac{2}{3w}\right)^{|G|} (\sqrt{3}v)^{|\partial G|}.
\end{equation}

We compare this result with the high-temperature expansion of the 3D Ising model in a uniform external magnetic field $h$, defined as $Z_{\mathrm{Ising}} = \sum_{\{\sigma_{\bm{r}}\}} \exp\left(K \sum_{\langle \bm{r}, \bm{r}' \rangle} \sigma_{\bm{r}} \sigma_{\bm{r}'} + h \sum_{\bm{r}} \sigma_{\bm{r}}\right)$. The expansion yields
\begin{equation}
    Z_{\mathrm{Ising}} = (\cosh K)^{2N} (\cosh h)^N 2^N \sum_{G} x^{|G|} y^{|\partial G|}.
\end{equation}
\rev{Matching these graph weights gives the approximate bare Ising parameters:}
\begin{align}
    x &\coloneqq \tanh K = \frac{2}{3w}, \\
    y &\coloneqq \tanh h = \sqrt{3}v.
\end{align}

\section{Magnetic Observables from Monte Carlo Simulations}
{%
To characterize the three regimes with conventional magnetic diagnostics, we measure---in addition to the specific heat, the macroscopic flux $\bm{P}$, and the monopole density of the main text---the $S^z=0$ occupation and its susceptibility,
\begin{equation}
m_0\coloneqq 1-\frac{1}{2N}\sum_\ell\langle(S_\ell^z)^2\rangle,\qquad
\chi_0\coloneqq \frac{1}{2NT}\bigg[\Big\langle\Big(\sum_\ell(S_\ell^z)^2\Big)^2\Big\rangle-\Big\langle\sum_\ell(S_\ell^z)^2\Big\rangle^2\bigg].
\end{equation}
\rev{Here $m_0$ is the uniform $S^z=0$ occupation and $\chi_0=\partial m_0/\partial\mu$ its susceptibility to the single-ion anisotropy. These are occupation observables, rather than the dual nematic order parameter $\langle e^{2i\theta_{\bm{r}}}\rangle$ and its susceptibility. In the strict monopole-free limit, $c=(\mu^2/T)\chi_0$, so their singular behavior is not an independent diagnostic.} The magnetic structure factor is built from the Cartesian moments $\bm{m}_\ell\coloneqq S_\ell^z\hat{\bm{e}}_\ell$ ($\hat{\bm{e}}_\ell$ the local $\langle111\rangle$ axis) as $S^{ab}(\bm{q})\coloneqq \frac{1}{2N}\langle m^a(\bm{q})m^b(-\bm{q})\rangle$. Its spin-flip channel $S(\bm{q})$~\cite{Fennell2009}---\rev{the polarization-resolved projection of transverse magnetic correlations measured in polarized neutron scattering}---carries the pinch points [shown in Fig.~\ref{fig:crossover}(e) of the main text].

\begin{figure}[t]
\centering
\includegraphics[width=0.6\linewidth]{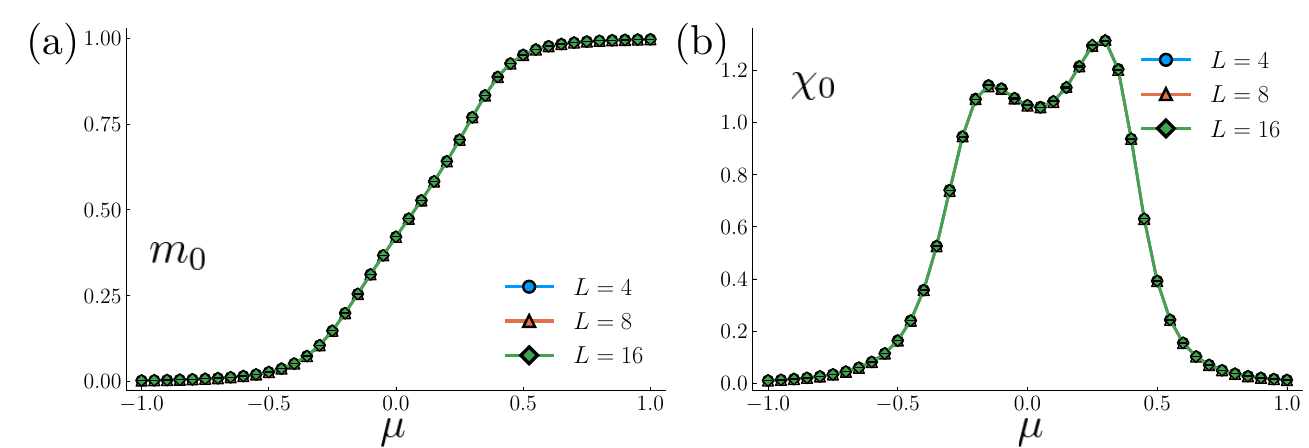}
\caption{(a) The $S^z=0$ occupation $m_0$ and (b) its susceptibility $\chi_0$, both versus the single-ion parameter $\mu$ at $T=0.3$ ($J=1$), for $L=4$, $8$, and $16$. The occupation $m_0$ evolves smoothly across the diagram, \rev{while $\chi_0$ has broad maxima in both crossover regions and little size dependence over the sizes shown.}}
\label{fig:magnetic}
\end{figure}

Figure~\ref{fig:magnetic} shows $m_0$ and $\chi_0$ versus $\mu$ at $T=0.3$. \rev{The occupation changes smoothly from a predominantly $S^z=\pm1$ population at negative $\mu$ to a predominantly $S^z=0$ population at positive $\mu$. The susceptibility has broad maxima in both crossover regions. Its weak size dependence over the displayed sizes is consistent with rounding, but neither the occupation susceptibility nor the specific heat alone establishes the critical universality class. The dual charge-$2$ characterization of the nematic regime is discussed in Sec.~I.}
}

\begingroup
\color{black}
\section{Quantitative specific-heat comparison and monopole-free control}
\label{sec:quantitative}

Figure~\ref{fig:quantitative} compares the $L=16$ data of Fig.~\ref{fig:crossover}(a) with dual calculations at $T=0.3$. The angular calculation samples the exact, unexpanded Eq.~\eqref{eq:angular-finite}, retaining all charges $q=1,\ldots,4$. Positive-weight sampling is restricted to $w<1/2$, or $\mu>T\ln2\simeq0.20794$, with a positive site factor for the parameters considered; at $v=0$ the site factor is unity. The negative-$\mu$ calculation samples the approximate graph sum in Eq.~\eqref{eq:graph-finite}, with weight $x^m y^n$, $x=2/(3w)$, $y=\sqrt3v$, where $m$ counts occupied edges and $n$ counts odd-degree vertices.

The specific heat per pyrochlore spin is $c=\beta^2(2N)^{-1}\partial_\beta^2\ln Z$, with $J,\mu$ fixed. Both fugacities vary with $\beta=1/T$ at finite $v$; only $w$ varies at strict $v=0$. Writing $\mathcal W$ for the full integrand of Eq.~\eqref{eq:angular-finite} gives the estimator, including its contact term,
\[
 c=\frac{\beta^2}{2N}\left[\operatorname{Var}(\partial_\beta\ln\mathcal W)+\langle\partial_\beta^2\ln\mathcal W\rangle\right].
\]
For Eq.~\eqref{eq:graph-finite},
\[
 c=\frac{\beta^2}{2N}\left[\mu^2\operatorname{Var}(m)+\frac{J^2}{4}\operatorname{Var}(n)-\mu J\operatorname{Cov}(m,n)\right],
\]
with the $n$ terms absent at $v=0$. Each point combines four replicas, with $2\times10^3$--$10^4$ equilibration sweeps, $1.6\times10^4$--$2.56\times10^5$ measurement sweeps, and $1000$--$2000$-sweep blocks. Error bars use the larger of the pooled block-jackknife standard error and the independent-replica scatter.

\begin{figure}[t]
\centering
\includegraphics[width=\linewidth]{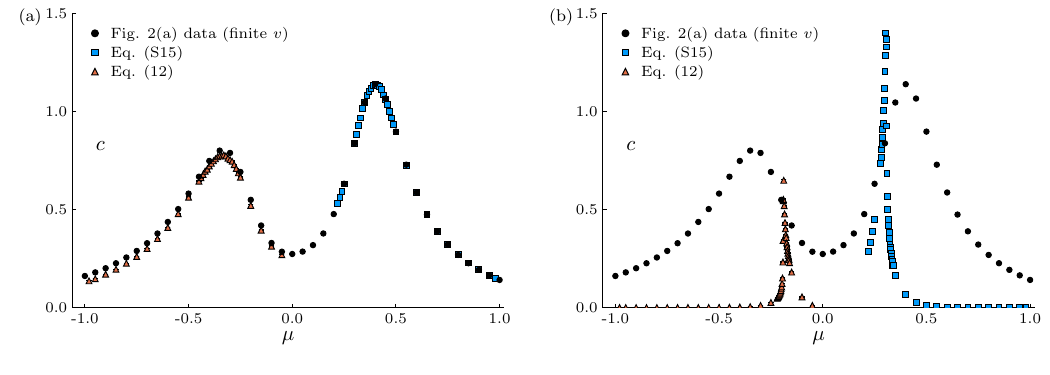}
\caption{\rev{Specific heat per pyrochlore spin at $T=0.3$, $L=16$. Squares use Eq.~\eqref{eq:angular-finite}; triangles use Eq.~\eqref{eq:graph-finite}, only for $\mu<0$. (a) Finite $v=e^{-J/(2T)}$ with $J=1$; (b) strict $v=0$, with $w=e^{-\mu/T}$ retained. Circles in both panels reproduce the same finite-$v$ data of Fig.~\ref{fig:crossover}(a). New points have one-standard-error bars; reference errors were unavailable.}}
\label{fig:quantitative}
\end{figure}

The angular calculation reproduces the finite-$v$, positive-$\mu$ peak without fitted parameters, providing an independent check of the exact representation. Eq.~\eqref{eq:graph-finite} reproduces the negative-$\mu$ peak height and full width at half maximum within about $3\%$ and $4\%$. These interpolation-based comparisons omit unavailable reference errors and interpolation uncertainties. Deviations may reflect the Pauling approximation to background-spin correlations, omission of $(3/2)^{V_4}$, and neglect of $Q=\pm2,\pm3,\pm4$ monopoles. Removing monopoles sharpens the peaks and shifts their positions. For comparison, the constrained critical values are $\mu_{c,\mathrm{Ising}}\simeq-0.1910$ and $\mu_{c,XY}\simeq0.3057$~\cite{Pandey2025}, while Eq.~\eqref{eq:graph-finite} predicts $\mu_c\simeq-0.1901$. Finite-size heat maxima need not coincide with these values; the negative-$\mu$ controls remain approximate.

Figure~\ref{fig:heat-peak-size} shows the largest measured heat values on common near-peak grids for $L=4,8,16,24,32$, with spacings $\Delta\mu=0.005$ at finite $v$ and $0.0005$ at $v=0$, plus boundary points. The $L=4$, $v=0$ graph shoulder is additionally extended. Finite-$v$ heights vary little, whereas both $v=0$ peaks grow over these sizes. Together with Fig.~\ref{fig:quantitative}, this supports monopole rounding. The $XY$-side growth is compatible with slow approach to a finite cusp ($\alpha<0$)~\cite{Campostrini2006}; saturation is unresolved. The heat data alone do not determine critical exponents or establish thermodynamic analyticity.

\makeatletter
\setlength{\@fpsep}{18pt}
\makeatother
\begin{figure}[t]
\centering
\includegraphics[width=\linewidth]{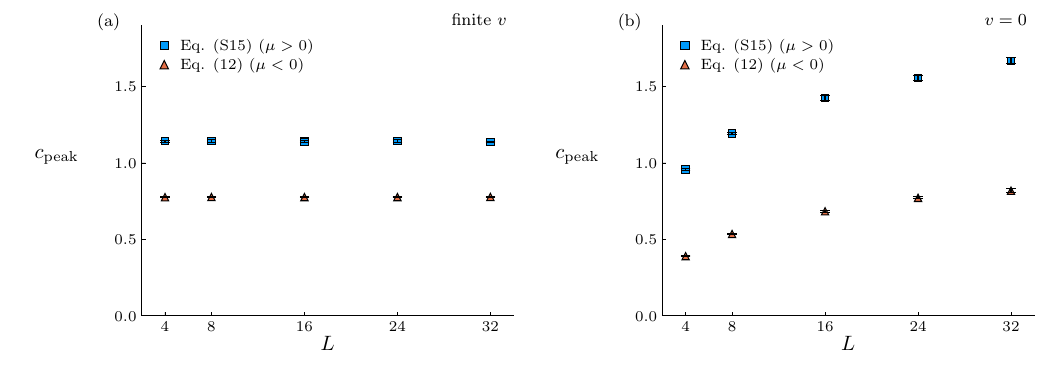}
\caption{\rev{Largest sampled specific heat per pyrochlore spin versus $L$ at $T=0.3$: (a) finite $v=e^{-J/(2T)}$ with $J=1$; (b) strict $v=0$. Squares use the exact Eq.~\eqref{eq:angular-finite} for the positive-$\mu$ peak; triangles use the approximate Eq.~\eqref{eq:graph-finite} for the negative-$\mu$ peak. Error bars are one standard error at the selected grid point and exclude grid and maximum-selection uncertainties; the points are not estimates of continuous-curve maxima.}}
\label{fig:heat-peak-size}
\end{figure}

\section{Finite-size scaling of the exact angular representation}
\label{sec:angular-scaling}

To examine criticality beyond heat-peak heights, we test the positive-$\mu$ critical point and its monopole-induced cutoff using Eq.~\eqref{eq:angular-finite} in the positive-weight domain. These simulations retain the full bond weight and all charge harmonics, so they test the original partition function in dual variables.

\subsection{Monopole-free critical point}

At $T=0.3$, $v=0$, four replicas at each of seven nearby $\mu$ values cover $L=16,24,28,32,36,40,48,56,64,72,80$. Each uses $8192$--$24576$ equilibration sweeps and $131072$ measurement sweeps, with $2048$-sweep blocks. A sweep combines $4$--$10$ cluster reflections with a local Metropolis pass.

For $N=8L^3$, define
\begin{equation}
 M=\sum_{\bm r}e^{i\theta_{\bm r}},\qquad S_0=\frac{\langle|M|^2\rangle}{N},\qquad U_4=\frac{\langle|M|^4\rangle}{\langle|M|^2\rangle^2}.
 \label{eq:zero-observables}
\end{equation}
With $S_k$ the structure factor averaged over $\bm k_\gamma=(2\pi/L)\hat{\bm e}_\gamma$, in cubic-cell units,
\begin{equation}
 \frac{\xi_\theta}{L}=\frac{1}{2\pi}\sqrt{\frac{S_0}{S_k}-1}.
 \label{eq:zero-length}
\end{equation}
At each crossing $R(L,\mu_\times)=R(2L,\mu_\times)$ for $R=U_4,\xi_\theta/L$, the size-doubling quotients~\cite{Ballesteros1996} are
\begin{align}
 \eta_{\mathrm{eff},R}(L)&=2-\frac{\ln[S_0(2L)/S_0(L)]}{\ln2},\nonumber\\
 y_{t,\mathrm{eff},R}(L)&=\frac{\ln[(\partial_\mu R)_{2L}/(\partial_\mu R)_L]}{\ln2},\qquad \nu_{\mathrm{eff},R}=y_{t,\mathrm{eff},R}^{-1}.
 \label{eq:quotients}
\end{align}
All quantities are evaluated at $\mu_\times$. Measured moments and derivatives are linearly interpolated; derivatives of parameter-independent angular moments use $\partial_\mu\langle O\rangle=\operatorname{Cov}(O,\partial_\mu\ln\mathcal W)$ and the chain rule, not interpolation slopes.

The six pairs are $16/32,24/48,28/56,32/64,36/72,40/80$. We fit
\begin{align}
 \eta_{\mathrm{eff},R}(L)&=\eta_\infty+A_R(L/16)^{-\omega},\nonumber\\
 y_{t,\mathrm{eff},R}(L)&=y_{t,\infty}+B_R(L/16)^{-\omega},\qquad \nu_\infty=y_{t,\infty}^{-1},
 \label{eq:quotient-correction}
\end{align}
with common limits, separate amplitudes, and fixed $\omega=0.789$~\cite{Hasenbusch2019}. Generalized least squares retains the covariance across pairs and crossing conditions. Errors use a joint replica/block bootstrap: $4720$ of $5000$ draws have unique crossings for all quotients, so intervals are conditional on this selection and the correction model.

Figure~\ref{fig:xy-correction} gives
\begin{equation}
 \eta_\infty=0.0351(48),\qquad \nu_\infty=0.666(16),
 \label{eq:xy-exponents}
\end{equation}
with $\chi^2/\mathrm{dof}=2.19/9$ and $6.65/9$, respectively. These agree with $\eta_{XY}=0.03810(8)$ and $\nu_{XY}=0.67169(7)$~\cite{Hasenbusch2019}, supporting $3$D $XY$ criticality. The statistical errors exclude correction-form uncertainty and residual initial-state dependence; two-replica initial-state subsets differ by $2.4\sigma$ in $\nu$, with smaller sensitivity at larger sizes. These limitations preclude a high-precision independent exponent determination. In particular, the sign of $\alpha=2-3\nu$ is unresolved; the finite-cusp interpretation uses the established $XY$ class.

\begin{figure}[t]
\centering
\includegraphics[width=\linewidth]{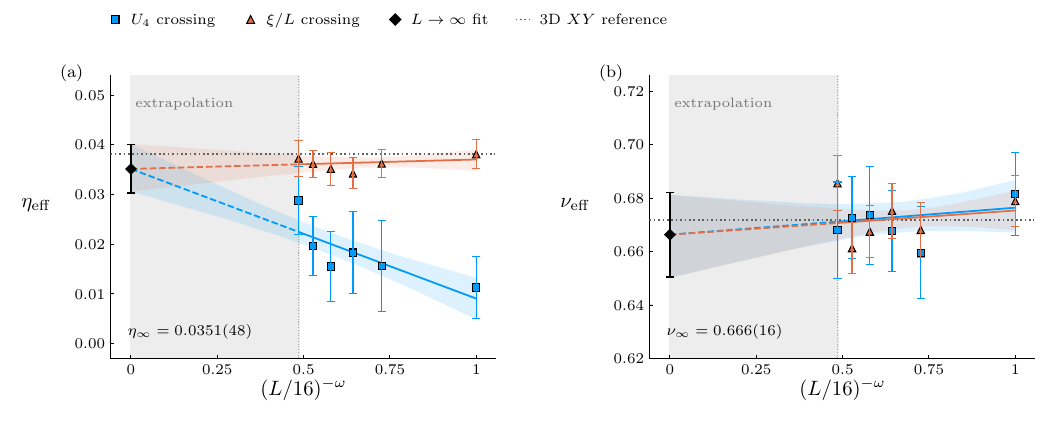}
\caption{\rev{Size-doubling estimates of (a) $\eta$ and (b) $\nu$ at $T=0.3$, $v=0$. Squares and triangles use $U_4$ and $\xi_\theta/L$ crossings. Fits use Eq.~\eqref{eq:quotient-correction} with fixed $\omega=0.789$; $L$ is the smaller size. Black diamonds are fitted common limits; dotted lines are reference $XY$ values. Solid and dashed curves cover measured and extrapolated ranges. Bars are conditional bootstrap standard deviations; bands are $16$--$84\%$ parameter intervals, excluding model systematics.}}
\label{fig:xy-correction}
\end{figure}

\subsection{Finite-monopole cutoff and field scaling}

At $T=0.3$, $\mu=0.305714$, we simulate $L=16,24,32,48$ and $v=10^{-4},10^{-3},10^{-2}$, varying $J=-2T\ln v$. Three further conditions match the nominal field-scaling coordinate introduced below across sizes. Each uses four replicas, $16384$ equilibration and $131072$ measurement sweeps, with $2048$-sweep blocks; local Metropolis updates are combined with generalized ghost-cluster reflections~\cite{KentDobias2018}.

Relative to the monopole-generated field, let $m_\parallel=\cos\theta$ and $m_\perp=\sin\theta$. For $M_\alpha(\bm k)=\sum_{\bm r}m_\alpha(\bm r)e^{i\bm k\cdot\bm r}$, define
\begin{align}
 S_\alpha^c(\bm k)&=\frac{\langle|M_\alpha(\bm k)|^2\rangle-|\langle M_\alpha(\bm k)\rangle|^2}{N},\nonumber\\
 \chi_\alpha&=S_\alpha^c(\bm0),\qquad
 \frac{\xi_\alpha}{L}=\frac{1}{2\pi}\sqrt{\frac{S_\alpha^c(\bm0)}{\frac13\sum_\gamma S_\alpha^c(\bm k_\gamma)}-1}.
 \label{eq:connected-angular}
\end{align}
These dimensionless susceptibilities and lengths describe connected dual-angle fluctuations, distinct from $\chi_0$ and spin screening. The $v=0$ references are interpolated to the same $\mu$, with $\chi_\alpha=S_0/2$.

Figure~\ref{fig:finite-v-cutoff} shows critical growth at $v=0$ and a finite-$v$ cutoff. At $v=10^{-3}$, $\chi_\perp\simeq138$ and $\xi_\perp\simeq4.0$ are nearly size independent for $L=24$--$48$; larger $v$ gives a shorter cutoff.

\begin{figure}[t]
\centering
\includegraphics[width=0.90\linewidth]{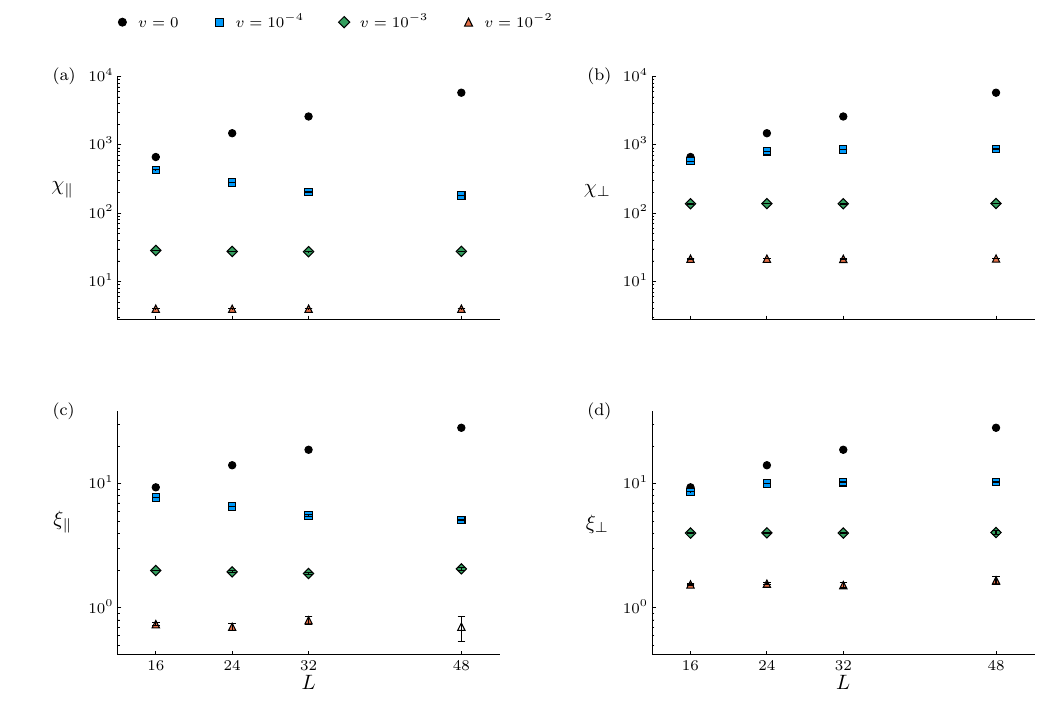}
\caption{\rev{Connected dual-angle (a),(b) susceptibilities and (c),(d) lengths at $T=0.3$, $\mu=0.305714$. Black circles are $v=0$ references. Bars use the larger of block/replica errors and bootstrap standard deviations. The open $L=48,v=10^{-2}$ point in (c) uses a conditional $16$--$84\%$ interval; $1.38\%$ of bootstrap draws have a nonpositive square-root argument. Parameter-location uncertainties are excluded.}}
\label{fig:finite-v-cutoff}
\end{figure}

Near the $XY$ critical point, the ordering field is $h=c_hv+O(v^3)$ with nonuniversal $c_h$~\cite{Powell2012}. With $t$ the thermal scaling field vanishing at the zero-field critical point, the leading finite-size forms are
\begin{equation}
 \chi_\alpha=L^{2-\eta}\mathcal X_\alpha(tL^{1/\nu},hL^{y_h}),\qquad
 \frac{\xi_\alpha}{L}=\mathcal R_\alpha(tL^{1/\nu},hL^{y_h}),\qquad y_h=\frac{5-\eta}{2}.
 \label{eq:field-scaling}
\end{equation}
At $t=0$, the thermodynamic field length scales as $\xi_h\propto v^{-1/y_h}\simeq v^{-0.40307}$. Figure~\ref{fig:finite-v-scaling} uses the nominal coordinate $x=(2v)L^{y_h}$ with reference $\eta=0.03810$, $y_h=2.48095$. Approximate agreement at matched $x$ supports ordering-field scaling; the exponents are fixed, and field metrics and nonlinear corrections remain undetermined.

\begin{figure}[t]
\centering
\includegraphics[width=0.90\linewidth]{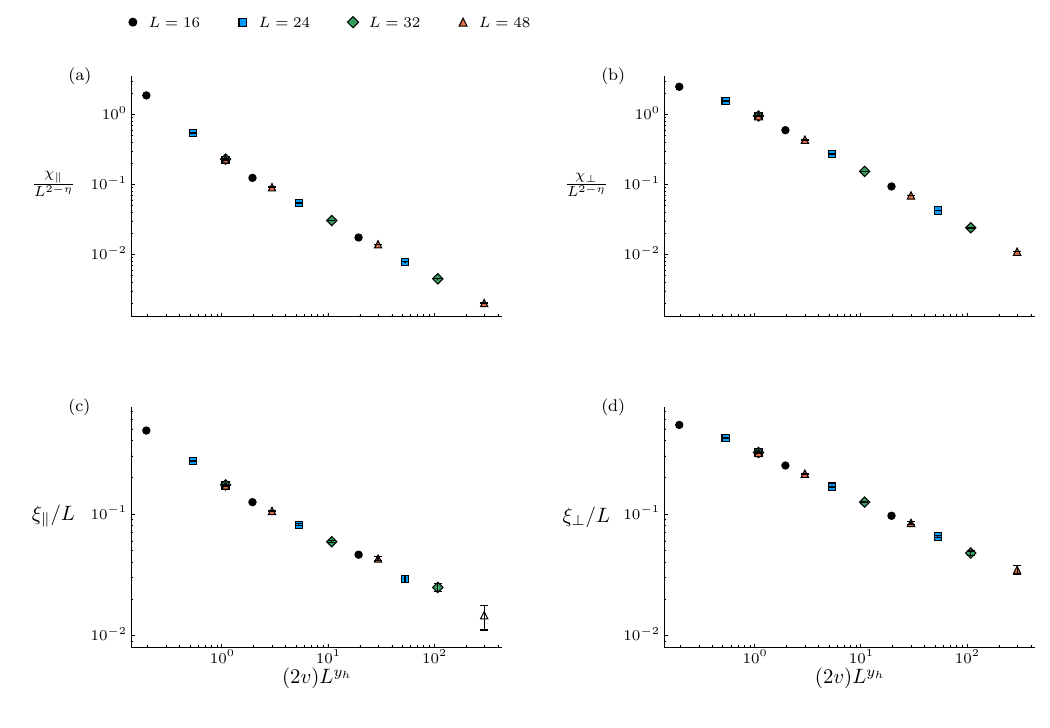}
\caption{\rev{Scaled connected susceptibilities and lengths versus $x=(2v)L^{y_h}$ for the fifteen finite-$v$ conditions in Fig.~\ref{fig:finite-v-cutoff}. Reference $XY$ exponents are fixed; four sizes share $x=1.08450569$. Errors and the open symbol follow Fig.~\ref{fig:finite-v-cutoff}.}}
\label{fig:finite-v-scaling}
\end{figure}

These results support the $XY$ description and its monopole-generated cutoff. They concern much smaller $v$ than the value $v\simeq0.189$ in Fig.~\ref{fig:crossover}, so they do not predict that heat peak quantitatively or test the bare coefficients of Eq.~\eqref{eq:xy-leading-finite}. The heat comparison in Sec.~\ref{sec:quantitative} addresses those finite-$v$ peaks; no global claim of analyticity follows from these finite-size tests.

\endgroup

\end{document}